\documentclass[english,conference]{IEEEtran}
\usepackage[T1]{fontenc}
\usepackage[latin9]{inputenc}
\usepackage{float}
\usepackage{amsmath}
\usepackage{amssymb}
\usepackage{graphicx}

\usepackage{mathrsfs}
\usepackage{amsfonts}
\usepackage{graphicx,cite,epsfig,amssymb,amsmath}
\usepackage{color,xcolor}
\usepackage{pifont}
\usepackage{stmaryrd}
\usepackage{setspace}
\usepackage{subfigure}
\usepackage{cite}
\usepackage{array}\usepackage{float}\usepackage{multirow}
\usepackage{algorithmic,algorithm}
\usepackage{subeqnarray}
\usepackage{cases}
\usepackage{url}
\providecommand{\tabularnewline}{\\}
\floatstyle{ruled}
\newfloat{algorithm}{tbp}{loa}
\providecommand{\algorithmname}{Algorithm}
\floatname{algorithm}{\protect\algorithmname}

\newtheorem{thm}{Theorem}
\makeatletter

\providecommand{\tabularnewline}{\\}
\floatstyle{ruled}
\newfloat{algorithm}{tbp}{loa}
\providecommand{\algorithmname}{Algorithm}
\floatname{algorithm}{\protect\algorithmname}

\usepackage{array}\usepackage{float}\usepackage{multirow}

\ifCLASSOPTIONcompsoc
\usepackage[caption=false,font=normalsize,labelfont=sf,textfont=sf]{subfig}
\else
\usepackage[caption=false,font=footnotesize]{subfig}
\fi

\providecommand{\tabularnewline}{\\}
\floatstyle{ruled}
\newfloat{algorithm}{tbp}{loa}
\providecommand{\algorithmname}{Algorithm}
\floatname{algorithm}{\protect\algorithmname}
\usepackage{cite}\@ifundefined{showcaptionsetup}{}{%
 \PassOptionsToPackage{caption=false}{subfig}}
\usepackage{subfig}
\IEEEoverridecommandlockouts
\makeatother

\usepackage{babel}
\begin{document}

\title{Energy- and Spectral-Efficiency Tradeoff in Full-Duplex Communications}

\author{\authorblockN{Dingzhu Wen\IEEEauthorrefmark{1}, Guanding Yu\IEEEauthorrefmark{1}, Rongpeng Li\IEEEauthorrefmark{2}, Yan Chen\IEEEauthorrefmark{2}, and Geoffrey Ye Li\IEEEauthorrefmark{3}}
\authorblockA{\IEEEauthorrefmark{1}College of ISEE, Zhejiang University, Hangzhou 310027, China}
\authorblockA{\IEEEauthorrefmark{2}Central Research Institute, Huawei Technologies Co., Ltd, Shanghai, China}
\authorblockA{\IEEEauthorrefmark{3}School of ECE, Georgia Institute of Technology, Atlanta, GA, USA}
}

\maketitle
\begin{abstract}
This paper investigates the tradeoff between \emph{energy-efficiency} (EE) and  \emph{spectral-efficiency} (SE) for \emph{full-duplex} (FD) enabled cellular networks. We assume that small cell base stations are working in the FD mode while user devices still work in the conventional \emph{half-duplex} (HD) mode. First, a necessary condition for a FD transceiver to achieve better EE-SE tradeoff than a HD one is derived. Then, we analyze the EE-SE relation of a FD transceiver in the scenario of single pair of users and obtain a closed-form expression. Next, we extend the result into the multi-user scenario and prove that EE is a quasi-concave function of SE in general and develop an optimal algorithm to achieve the maximum EE based on the Lagrange dual method. Our analysis is finally verified by extensive numerical results.
\end{abstract}

\section{Introduction}
\emph{Full-duplex} (FD) communications have achieved more and more attentions because of the potential to double the wireless link capacity\cite{FD_Challenges_Opportunities}. With the current \emph{self-interference} (SI) cancellation techniques, such as propagation domain suppression, analog cancellation, and digital cancellation, SI can be mitigated to a sufficiently low level, to make the FD communications practically implementable. Recently, there has been some work in facilitating the application of FD enabled cellular networks from the perspective of enhancing \emph{spectral-efficiency} (SE) \cite{FD_EE,FD_SE}.

Since power consumption of mobile devices is increasing rapidly whereas the battery capacity is still limited, \emph{energy-efficiency} (EE) becomes a more and more important metric for cellular networks. The EE design of wireless systems has been investigated since a decade ago. It is well known that the EE and SE cannot be simultaneously optimized in general, especially when the circuit power consumption is considered. The EE-SE tradeoff in downlink OFDMA networks has been initially studied in \cite{OFDMA_tradeoff}. For OFDMA networks, EE is a quasi-concave function of SE. The quasi-concavity of the EE-SE relation has also been extended into other systems, such as the amplify-and-forward relay network \cite{EESE_Huang}, the type-I ARQ system \cite{EESE_Wu}, and the cognitive radio network \cite{EESE_Zhang}. However, the EE-SE tradeoff of the FD enabled network has not been addressed yet.

In this paper, we will investigate the EE-SE relation in FD enabled networks where small cell \emph{base stations} (BSs) are working in the FD mode while user devices work in the conventional \emph{half-duplex} (HD) mode. The main challenge here is to deal with both SI and \emph{co-channel interference} (CCI), which render it difficult to analyze the EE-SE tradeoff. We first find a necessary condition for FD communications to outperform the conventional HD communications in term of EE-SE tradeoff when the \emph{residual SI} (RSI) power is constant as in \cite{fixed_RSI}. Then, the closed-form expression of the EE-SE tradeoff is found and the EE is proved to be a quasi-concave function of the SE when considering only one pair of users. Moreover, the quasi-concavity is further extended to the multi-user scenario. Based on the quasi-concavity, the optimal algorithm is developed to achieve the maximum EE for a given SE region, which contains two loops. The inner loop solves the EE maximization problem for a given SE by the \emph{Lagrangian dual decomposition} (LDD) technique\cite{convex_optimization} while the outer loop achieves the optimal EE for different SEs based on the bisection method.

The rest of the paper is organized as follows. In Section II, we describe the system model for the FD network and formulate the EE-SE tradeoff problem. In Section III, the necessary condition for FD communications to outperform HD communications in term of EE-SE tradeoff is derived. In Section IV, we analyze the EE-SE relation for a single pair of users. In Section V, we extend our study to the multi-user scenario and propose an optimal algorithm to maximize EE for a given SE region. Numerical results are presented in Section VI and the whole paper is concluded in Section VII.

\section{System Model and Problem Formulation}
In this section, we introduce the system model and then formulate the EE-SE tradeoff problem.
\subsection{System model}
\begin{figure}[htp]
\center
\includegraphics[width=0.32\textwidth]{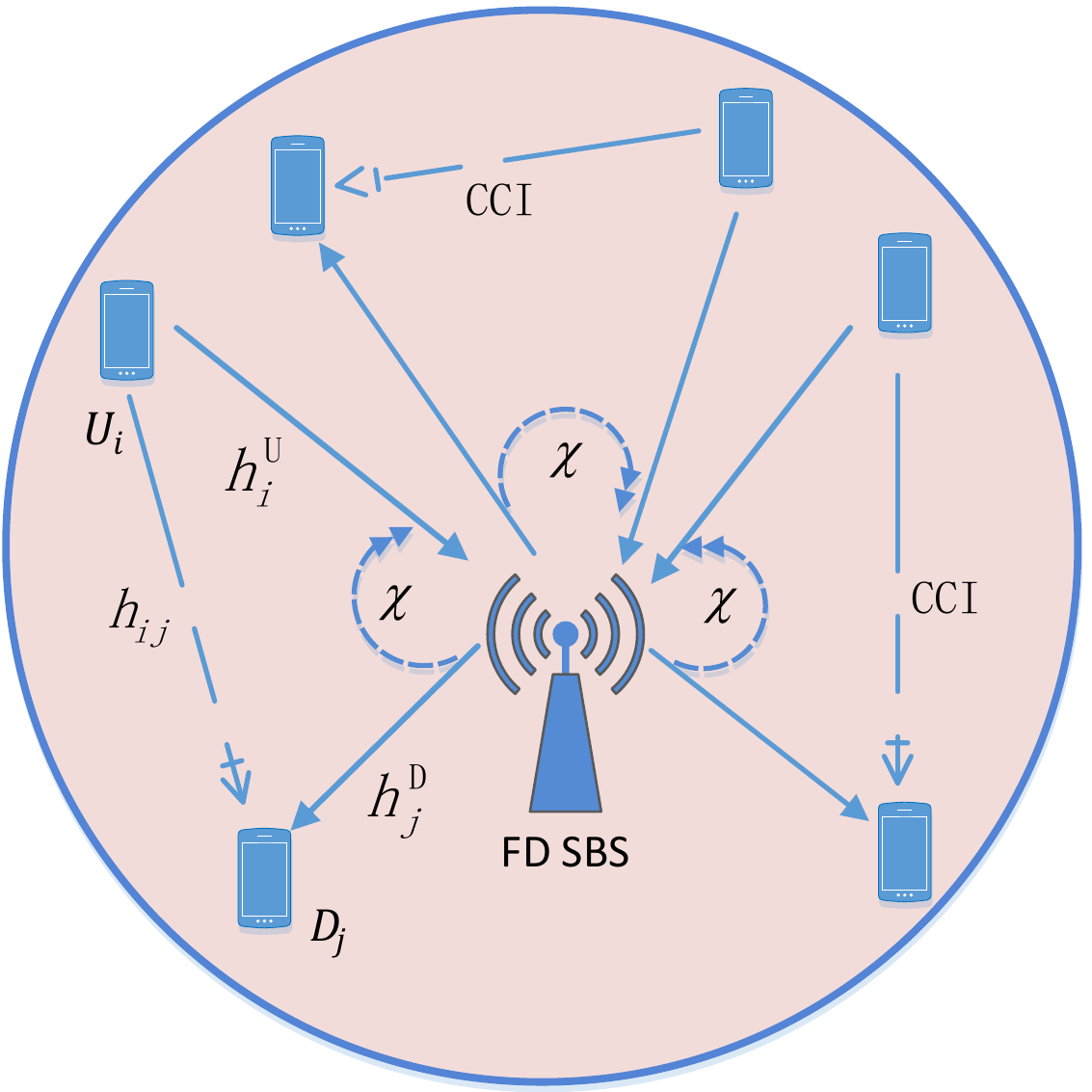}
\caption{The system model of the FD network.}\label{sysmodel}
\end{figure}

As depicted in Fig. \ref{sysmodel}, we consider a \emph{time-division} (TD) single picocell network with the \emph{small BS} (SBS) located at the center of the cell. The SBS is FD enabled whereas the user devices still work in the conventional HD mode due to their limited hardware capability of SI cancellation.

When the SBS is in FD mode, an uplink user and a downlink user can be paired to communicate simultaneously. At the same time, two kinds of interference are consequently incurred: SI at the SBS affecting the uplink transmission and CCI affecting the downlink transmission. In this paper, we assume that the RSI power after SI cancellation is a constant, which can be known to the SBS in advance \cite{fixed_RSI}.
As in Fig. \ref{sysmodel}, denote $h_i^{\rm{U}}$ as the uplink \emph{carrier-to-noise ratio} (CNR) of user $i$, which can be expressed as
\begin{equation}
h_{i}^{\rm{U}}=G\beta_{i}d_{i}^{-\alpha}/N_0,
\end{equation}
where $G$ is the path loss constant, $\beta_{i}$ is the fading coefficient, $d_i$ is the distance from the user $i$ to the SBS, $\alpha$ is the path loss exponent, and $N_0$ is the power of the \emph{additive white Gaussian noise} (AWGN). In the sequel, for notation simplicity and without loss of generality, we assume that $N_0$ is normalized according to the transmit power, i.e., $N_0 = 1$. Similarly, we can define $h_{j}^{\rm{D}}$ as the downlink CNR of user $j$ and $h_{i,j}$ as the CCI CNR from uplink user $i$ to downlink user $j$. We assume that the SBS can acquire the instantaneous \emph{channel state information} (CSI) of all involved links to perform centralized resource allocation.

Consider a picocell with $M$ pairs of users. As in Fig. \ref{sysmodel}, uplink user $i$ and downlink user $j$ are assumed to be paired, i.e., they communicate simultaneously. In this case,  the achievable uplink and downlink data rates can be expressed as
\begin{equation}
\left\{
\begin{split}
&R_{ij}^{\rm{U}} = W\log_2(1+\dfrac{p_{ij}^{\rm{U}}h_i^{\rm{U}}}{1 + \chi}),\\
&R_{ij}^{\rm{D}} = W\log_2(1+\dfrac{p_{ij}^{\rm{D}}h_j^{\rm{D}}}{1 + p_{ij}^{\rm{U}}h_{ij} }),
\end{split}
\right.
\end{equation}
respectively, where $W$ is the system bandwidth, $p_{ij}^{\rm{U}}$ and $p_{ij}^{\rm{D}}$ are the uplink and downlink transmit powers, respectively, and $\chi$ is the normalized RSI power.

Denote $R_{\rm{tot}}$ as the system throughput, which can be expressed as
\begin{equation}
R_{\rm{tot}}=\sum_i\sum_j\gamma_{ij}(R_{ij}^{\rm{U}}+R_{ij}^{\rm{D}}),
\end{equation}
where $\gamma_{ij}$ is the normalized slot length allocated to user pair $(i,j)$. Denote $P_{\rm{tot}}$ as the total transmit power, as
\begin{equation}
P_{\rm{tot}}=\sum_{i=1}^M\sum_{j=1}^N\gamma_{ij}(p_{{ij}}^{\rm{U}}+p_{{ij}}^{\rm{D}}).
\end{equation}

\subsection{Problem formulation}
For the network mentioned above, the EE and SE can be defined as
\begin{equation}
\beta_{\rm{EE}}=\dfrac{R_{\rm{tot}}}{\omega P_{\rm{tot}}+P_{\rm{fix}}}~~~\text{and}~~~~\beta_{\rm{SE}}=\dfrac{R_{\rm{tot}}}{W},
\end{equation}
respectively, where $\omega$ represents the inverse of the power amplifier efficiency and $P_{\rm{fix}}$ is the total fixed circuit power consumption of the system.

Therefore, in this network, the EE-SE tradeoff problem can be formulated as maximizing the EE for a given SE. It can be mathematically formulated as
\begin{eqnarray*}
~~~~~~~~
\begin{array}{c}
\beta_{\rm{EE}}^{*}(R_{\rm{tot}})=\max\limits_{R_{ij}^{\rm{U}}, R_{ij}^{\rm{D}}, \gamma_{ij}} ~\dfrac{R_{\rm{tot}}}{\omega P_{\rm{tot}}+P_{\rm{fix}}},
\end{array}~~~~~~~~~~~~\eqref{eq:1}
\end{eqnarray*}
\vspace{-2.5em}
\begin{subequations}\label{eq:1}
subject to\\
\begin{align}
&\mathop{\sum\limits_i\sum\limits_j} \gamma_{ij}\leq 1, \label{eq:1.1}\\
&\sum_i\sum_j\gamma_{ij}(R_{{ij}}^{\rm{U}}+R_{{ij}}^{\rm{D}})= R_{\rm{tot}},\label{eq:1.2}\\
&\sum\limits_j \gamma_{ij}\geq \gamma_{\min}^{\rm{U}},~\forall i,\label{eq:1.3}\\
&\sum\limits_i \gamma_{ij} \geq \gamma_{\min}^{\rm{D}},~\forall j,\label{eq:1.4}
\end{align}
\end{subequations}
where \eqref{eq:1.3} and \eqref{eq:1.4} are the fairness constraints, which guarantee minimum amount of time-slots for both uplink and downlink users, respectively.

In the following sections, we approach the problem in \eqref{eq:1} in the following three aspects. First, the EE-SE tradeoff comparison between FD communications and HD communications is analyzed and a necessary condition for FD communications to be better than HD communications is derived. Then, the EE-SE tradeoff problem in the scenario of single pair of users is investigated to get some insights. In this scenario, EE is proved to be a quasi-concave function of SE. Next, we will show that EE is also quasi-concave on SE even in the multi-user scenario and hence a global optimal solution can be developed.

\section{FD or HD communications?}
Before investigating the EE-SE tradeoff problem for FD enabled networks, it is important to understand in which case FD communications are better than HD communications. In this section, a necessary condition for FD communications to be better than HD communications is found.

Denote $P_{\rm{F}}$ and $P_{\rm{H}}$ as the transmit powers of FD mode and HD mode, respectively. Denote $R_{\rm{F}}$ and $R_{\rm{H}}$ as the data rate of FD mode and HD mode, respectively. Then, we have
\begin{equation}
\left\{
\begin{split}
&R_{\rm{F}}=R_{ij}^{\rm{U}}+R_{ij}^{\rm{D}},\\
&R_{\rm{H}}=\max\big(\log_2(1+p_{\rm{H}}h_i^{\rm{U}}), \log_2(1+p_{\rm{H}}h_j^{\rm{D}})\big).
\end{split}
\right.
\end{equation}

The EE comparison problem between FD communications and HD communications can be formulated as comparing the EE with the same SE, that is, compare
\begin{equation}
\beta_{\rm{F}}=\dfrac{R_F}{\omega P_{\rm{F}}+P_{\rm{fix}}}~~~\text{and}~~~\beta_{\rm{H}}=\dfrac{R_H}{\omega P_{\rm{H}}+P_{\rm{fix}}},
\end{equation}
when $R_{\rm{F}}=R_{\rm{H}}=R_{\rm{tot}}$.

By comparing $\beta_{\rm{F}}$ and $\beta_{\rm{H}}$ for same SE, a necessary condition can be derived, as presented in the following theorem and proved in Appendix A.
\begin{thm}
If FD communications have a better EE-SE tradeoff than HD communications, the following condition must be satisfied.
\begin{equation}\label{eq:th1}
h_{ij}(1+\chi)<\min(h_i^{\rm{U}},h_j^{\rm{D}}).
\end{equation}
\end{thm}

From Theorem 1, we have the following intuitive but insightful observations.
\begin{itemize}
\item If the CNR of the CCI link, $h_{ij}$, is greater than a threshold, i.e., $(1+\chi)^{-1}\min(h_i^{\rm{U}},h_j^{\rm{D}})$, HD communications will have a better EE-SE tradeoff than FD communications.
\item The FD communications would have a better EE-SE tradeoff than the HD communications only if $\chi$ is less than a threshold, $h_{ij}^{-1}\min(h_i^{\rm{U}},h_j^{\rm{D}})-1$. Otherwise, HD mode should be used due to the large RSI power.
\end{itemize}

Note that although \eqref{eq:th1} in Theorem 1 is not a sufficient condition, it is close to the sufficient condition especially when the uplink CNR is close to the downlink CNR and the transmit power is small, as explained in Appendix A. Nevertheless, \eqref{eq:th1} will be used as the pre-condition where the FD mode should be used in the sequel.


\section{One Pair of Users}
In this section, we start with the scenario of single pair of users to gain some insights of the EE-SE tradeoff. In this case, transmit power allocation can be expressed in a closed-form. Moreover, the maximum EE, $\beta_{\rm{EE}}^{*}$, in this case can be proved to be a quasi-concave function of SE, and therefore the global optimal EE can be achieved for any given SE region.

For a given user pair $(i,j)$, the EE-SE tradeoff problem in \eqref{eq:1} can be simplified into
\begin{eqnarray*}
~~~~~~~~~~~~
\begin{array}{c}
\beta_{\rm{EE}}^{*}(R_{\rm{tot}}) = \max\limits_{R_{ij}^{\rm{U}},R_{ij}^{\rm{D}}}~~\dfrac{R_{\rm{tot}}}{\omega P_{\rm{tot}}+P_{\rm{fix}}},
\end{array}~~~~~~~~~
\eqref{eq:3}
\end{eqnarray*}
\vspace{-2.6em}
\begin{subequations}\label{eq:3}
subject to\\
\begin{align}
&R_{ij}^{\rm{U}}+R_{ij}^{\rm{D}}= R_{\rm{tot}}.\label{eq:3.1}
\end{align}
\end{subequations}

Obviously, for a given $R_{\rm{tot}}$, the solution to \eqref{eq:3} remains the same when the objective function is replaced by $P_{ij}^{\min} = \min (p_{ij}^{\rm{U}}+p_{ij}^{\rm{D}})$, since the other parameters are constants. Therefore, we first analyze the minimum transmit  power $P_{ij}^{\min}$ and then investigate the relation between $\beta_{\rm{EE}}^{*}(R_{\rm{tot}})$ and $R_{\rm{tot}}$.

By applying some simple mathematical derivations, the constraint in \eqref{eq:3.1} can be rewritten into
\begin{equation}\label{eq:4}
\begin{split}
&{p_{ij}^{\rm{U}}}^2h_i^{\rm{U}}h_{ij}+p_{ij}^{\rm{U}}p_{ij}^{\rm{D}}h_i^{\rm{U}}h_j^{\rm{D}}+p_{ij}^{\rm{U}}[(1-A)(1+\chi)h_{ij} \\
&+h_i^{\rm{U}}]+(1+\chi)p_{ij}^{\rm{D}}h_j^{\rm{D}}+(1-A)(1+\chi)=0,
\end{split}
\end{equation}
where $A=2^{R_{\rm{tot}}}$. Furthermore, the relation between $p_{ij}^{\rm{U}}$ and $p_{ij}^{\rm{D}}$ in \eqref{eq:4} forms a conic curve with the solution to $P_{ij}^{\min}$ locating at $\dfrac{{\rm{d}} p_{ij}^{\rm{D}}}{{\rm{d}} p_{ij}^{\rm{U}}}=-1$. By jointly considering \eqref{eq:4} and the first derivative, $P_{ij}^{\min}$ can be solved in a closed-form, as
\begin{equation}\label{eq:6}
P_{ij}^{\min}=\left\{
\begin{split}
&P_1,~\text{if}~C_1,\\
&P_2,~\text{if}~C_2,\\
&P_3, ~\text{if}~\overline{C_1}\&\overline{C_2},
\end{split}
\right.
\end{equation}
where the conditions $C_1$ and $C_2$ are defined as
\begin{equation}
\left\{
\begin{split}
&C_1: A\leq \dfrac{(h_j^{\rm{D}}-h_{ij})(1+\chi)}{h_i^{\rm{U}}-(1+\chi)h_{ij} },\\
&C_2: A\leq \dfrac{h_i^{\rm{U}}-(1+\chi)h_{ij} }{(h_j^{\rm{D}}-h_{ij})(1+\chi)},
\end{split}
\right.
\end{equation}
and $P_1$, $P_2$, and $P_3$ are defined as
\begin{equation}\label{eq:4-1}
\left\{
\begin{split}
&P_1=\dfrac{A-1}{h_j^{\rm{D}}},\\
&P_2=\dfrac{(A-1)(1+\chi)}{h_i^{\rm{U}}},\\
&P_3=\dfrac{2\sqrt{A(1+\chi)(h_i^{\rm{U}}-(1+\chi)h_{ij})(h_j^{\rm{D}}-h_{ij} )}}{ h_i^{\rm{U}}h_j^{\rm{D}}}\\
&~~~~~~~~+\dfrac{(1+A)(1+\chi)h_{ij}-h_i^{\rm{U}}-(1+\chi)h_j^{\rm{D}}}{ h_i^{\rm{U}}h_j^{\rm{D}}}.
\end{split}
\right.
\end{equation}

Furthermore, it can also be proved that the solution space of $(p_{ij}^{\rm{U}}, p_{ij}^{\rm{D}})$ is  $C_1\cup (\overline{C_1}\& \overline{C_2})$ or $C_2\cup (\overline{C_1}\& \overline{C_2})$.

From \eqref{eq:6}, we can further prove that $P_{ij}^{\min}$ is a monotone convex and strictly increasing function of $R_{\rm{tot}}$, as presented in the following theorem and proved in Appendix B.
\begin{thm}
Under the condition that FD communications are better than HD communications, that is, \eqref{eq:th1} holds, the minimum transmit power, $P_{ij}^{\min}$, increases with $R$ and is monotone convex.
\end{thm}

According to Theorem 2, we now come up with the following theorem, as proved in Appendix C.
\begin{thm}
The maximum EE, $\beta_{\rm{EE}}^{*}(R_{\rm{tot}})$ is a quasi-concave function of the SE, $\beta_{\rm{SE}}$, in the scenario of single pair of users.
\end{thm}

In \cite{OFDMA_tradeoff}, the quasi-concavity of EE-SE relation has also been demonstrated in downlink OFDMA networks with HD communications. However, different from the conventional HD networks, the FD network has two additional kinds of interference: SI and CCI. Besides, the uplink transmission and the downlink transmission are coupled, which makes the EE-SE tradeoff problem in FD networks more complicated than in HD networks. Theorem 3 shows that, under such complicate cases, the EE is still a quasi-concave function of the SE in the scenario of single pair of users. Therefore, the optimal EE can be achieved for any given SE region. In the next section, we will extend the result into the multi-user scenario.

\section{Multiple User Pairs}
In this section, we investigate the EE-SE tradeoff in the multi-user scenario. We also show that the EE is a quasi-concave function of the SE in this case. Based on the quasi-convexity, the global optimal algorithm to achieve the maximum EE for a given SE region will be developed.

Similar to Section IV, we first consider maximizing the EE for a given SE. In the multi-user scenario, the objective function of \eqref{eq:1} can be equivalently expressed as
\begin{equation}\label{eq:7}
P_{\rm{tot}}^{\min}=\min~\sum_{i=1}^M\sum_{j=1}^M\gamma_{ij}P_{ij},
\end{equation}
where $P_{ij}=p_{{ij}}^{\rm{U}}+p_{{ij}}^{\rm{D}}$ is the minimum total transmit power of user pair $(i,j)$ in FD mode.

Given the data rate of user pair $(i,j)$ as $R_{ij}=R_{{ij}}^{\rm{U}}+R_{{ij}}^{\rm{D}}$, we have $P_{ij}=P_{ij}^{\min}(R_{ij})$, which can be expressed in \eqref{eq:6}. Substituting $P_{ij}$ into \eqref{eq:7} leads to a non-convex optimization problem due to the non-convexity of $\gamma_{ij}P_{ij}$ and $\gamma_{ij}R_{ij}$. In the following, we will first transform the problem into a convex one.

We define an auxiliary variable as
\begin{equation}\label{eq:9}
\hat{R}_{ij}=\gamma_{ij}R_{ij}.
\end{equation}
Then, by substituting it into \eqref{eq:7}, the EE maximization problem can be transformed into
\begin{eqnarray*}
~~~~~~~~~~~~~~
\begin{array}{c}
P_{\rm{tot}}^{\min}=\min\limits_{\gamma_{ij},\hat{R}_{ij}}~~ \sum_i\sum_j\gamma_{ij}P_{ij}
\end{array}~~~~~~~~~~~~~~\eqref{eq:10}
\end{eqnarray*}
\vspace{-2.5em}
\begin{subequations}\label{eq:10}
subject to\\
\begin{align}
&\mathop{\sum\limits_i}\sum\limits_j \gamma_{ij}\leq 1, \label{eq:10.1}\\
&\sum_i\sum_j \hat{R}_{ij}= R_{\rm{tot}},\label{eq:10.2}\\
&\sum\limits_j \gamma_{ij}\geq \gamma_{\min}^{\rm{U}},~\forall i,\label{eq:10.3}\\
&\sum\limits_i \gamma_{ij} \geq \gamma_{\min}^{\rm{D}},~\forall j,\label{eq:10.4}
\end{align}
\end{subequations}
which is convex with regard to the variables ${\bf{\Gamma}}=\{\gamma_{ij}\}_{M\times N}$ and ${\bf{\hat{R}}}=\{\hat{R}_{ij}\}_{M\times N}$ when $\gamma_{ij}\in(0,1]$, as proved in Appendix D.

Based on the convexity of the EE maximization problem in \eqref{eq:10}, we further show that the EE is a quasi-concave function of the SE, as presented in the following theorem.
\begin{thm}
The optimal EE, $\beta_{\rm{EE}}^{*}(R_{\rm{tot}})$, is a quasi-concave function of the SE, $\beta_{\rm{SE}}=\dfrac{R_{\rm{tot}}}{W}$, in the scenario of multiple user pairs.
\end{thm}

The proof is similar to that of Theorem 3. Therefore, the global optimum EE, $\beta_{\rm{EE}}^{**}=\max\limits_{R_{\rm{tot}}}~ \beta_{\rm{EE}}^{*}(R_{\rm{tot}})$, can be achieved for any given SE region, as elaborated in the following.
\begin{itemize}
\item \emph{The inner loop:} For a fixed SE, i.e, $R_{\rm{tot}}$, solve the convex problem in \eqref{eq:10} to obtain $\beta_{\rm{EE}}^{*}(R_{\rm{tot}})$.
\item \emph{The outer loop:} Find the global optimum EE, $\beta_{\rm{EE}}^{**}$, within the SE region.
\end{itemize}

The detailed approach is omitted due to page limits.
\section{Numerical Results}
In the simulation, we consider a single picocell network with a radius of $150$ m. The FD enabled SBS is located at the center of the cell. User devices are uniformly distributed in the cell. The parameters of path loss fading and shadow standard deviation are according to \cite{simul_para}. Other major simulation parameters are listed in Table I.
\begin{table}[!htp]
\caption{Simulation parameters}
\vspace{-2em}
\centering{}%
\small
\begin{tabular}[t]{|l|l|}
\hline
Parameter & Value\tabularnewline
\hline
Cell radius, $r$ & ${\rm{150~m}}$\tabularnewline
\hline
Bandwidth, {\it{W}} &${\rm{10~ MHz}}$\tabularnewline
\hline
Noise power density &${\rm{-174~ dBm/Hz}}$\tabularnewline
\hline
Inverse of the power amplifier, $\omega$&1\tabularnewline
\hline
Path loss between user and SBS  &${\rm{145.4+37.5\log\left(d({\rm{\mathrm{km}}})\right)}}$\tabularnewline
\hline
Shadow standard deviation  &$10~{\rm{dB}}$\tabularnewline
\hline
Path loss between users & ${\rm{175.78+40\log\left(d({\rm{\mathrm{km}}})\right)}}$\tabularnewline
\hline
The circuit power consumption, $P_{\rm{fix}}$& 0.1 W\tabularnewline
\hline
\end{tabular}
\end{table}

\begin{figure}[!htp]
\centering
\begin{minipage}[b]{0.35\textwidth}
\subfigure[The minimum total transmit power, $P_{\rm{tot}}^{\min}$, with the SE, $\beta_{\rm{SE}}$.]{\includegraphics[width=\textwidth]{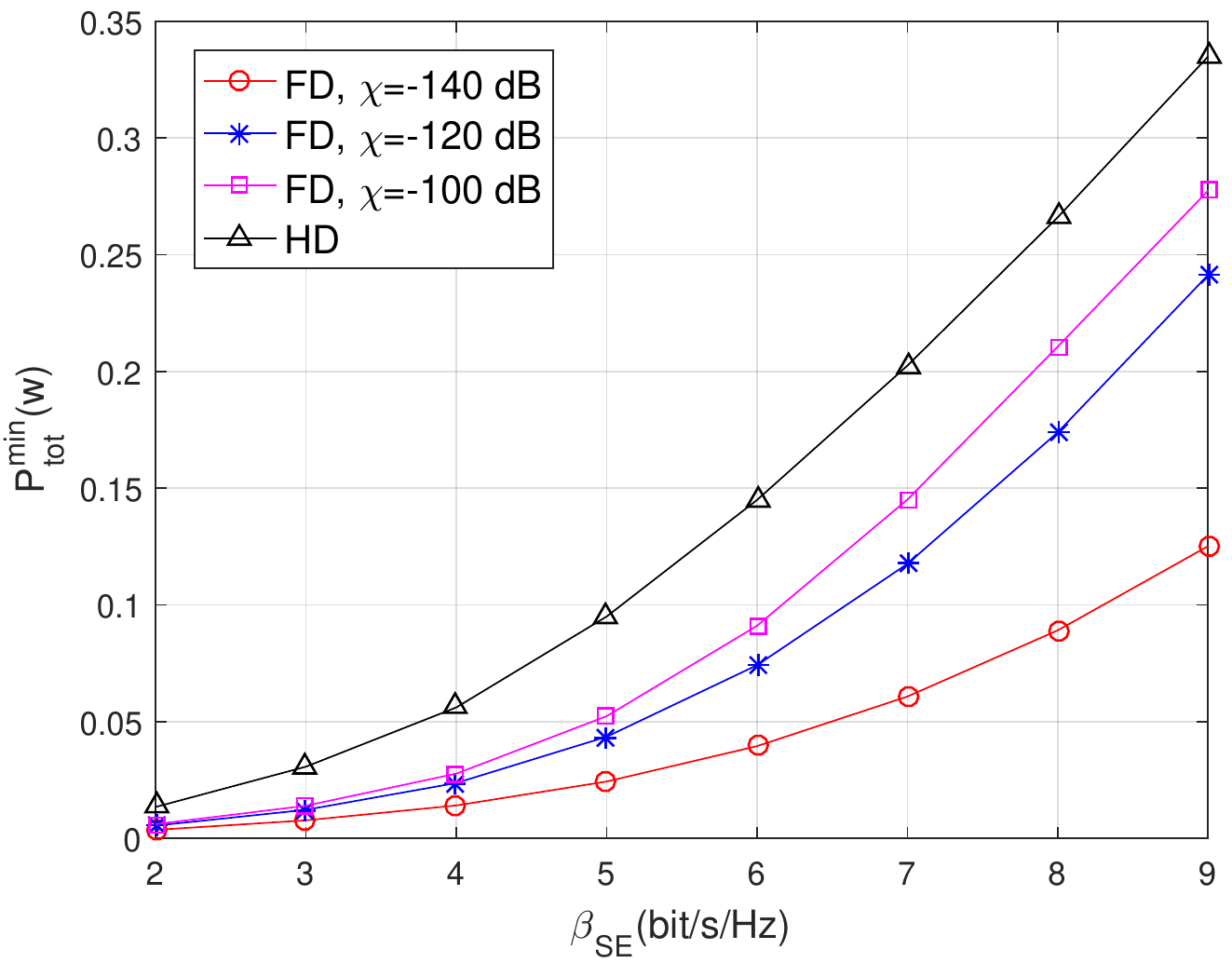}}
\hfill
\subfigure[The maximum EE, $\beta_{\rm{EE}}^{*}$, with the SE, $\beta_{\rm{SE}}$.]{\includegraphics[width=\textwidth]{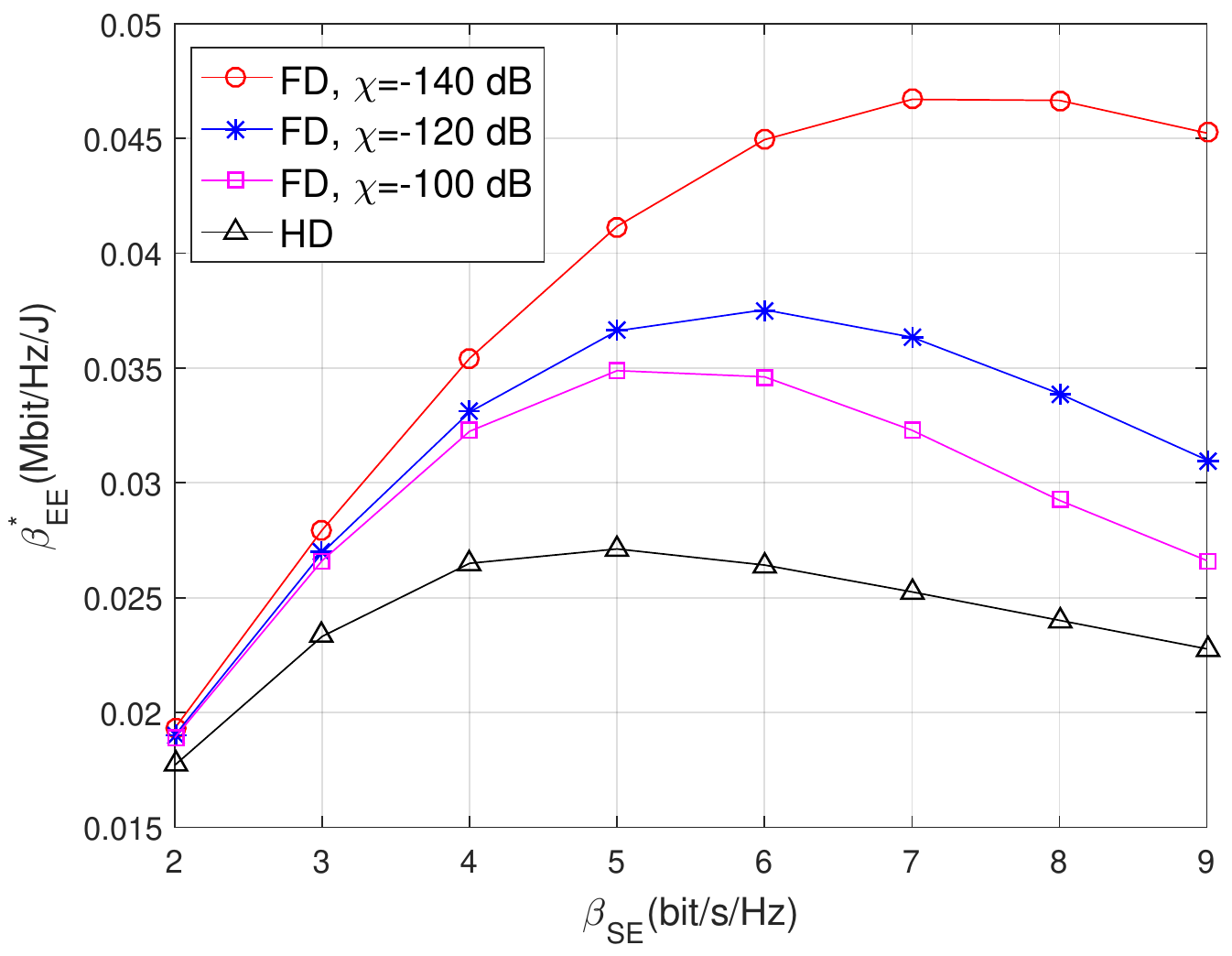}}
\caption{The relation of the maximum EE, $\beta_{\rm{EE}}^{*}$, and the SE, $\beta_{\rm{SE}}$. $M=6$.}\label{RPE_chi}
\end{minipage}
\end{figure}

Fig. \ref{RPE_chi} presents the relation of the maximum EE, $\beta_{\rm{EE}}^{*}(R_{\rm{tot}})$, and SE, $\beta_{\rm{SE}}$, when $M=6$. From the figure, $P_{\rm{tot}}^{\min}$ increases with the $\beta_{\rm{SE}}$, in both FD and HD networks. For the FD network, the larger the RSI power $\chi$ is, the more rapidly the transmit power increases. This is because that more transmit power is needed to surpass the RSI power to achieve the same SE when $\chi$ is large. In Fig. \ref{RPE_chi}(b), in both FD mode and HD mode, $\beta_{\rm{EE}}^{*}$ first increases and then decreases with the SE, which indicates that $\beta_{\rm{EE}}^{*}$ is a quasi-concave function of $\beta_{\rm{SE}}$, and validates Theorem 4.

\begin{figure}[!htp]
\center
\includegraphics[width=0.35\textwidth]{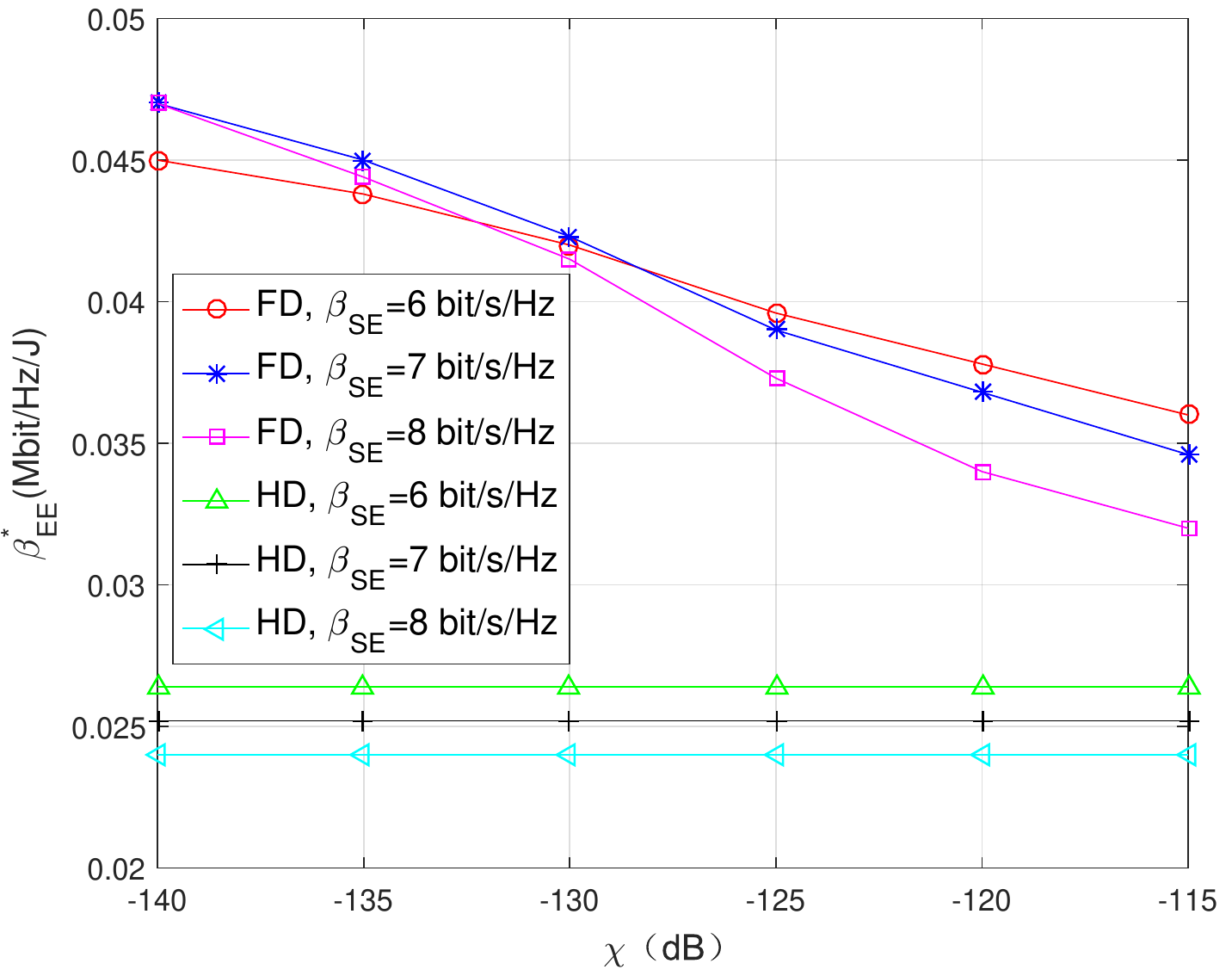}
\caption{EE-SE tradeoff for different RSI powers. $\beta_{\rm{SE}}=8~{\rm{bit/s/Hz}}$, $M=6$. }\label{Chi}
\end{figure}

Fig. \ref{Chi} depicts the EE performance with different RSI powers. Form the figure, $\beta_{\rm{EE}}^{*}(R_{\rm{tot}})$ decreases with the RSI power for a fixed SE in the FD network. The larger the required SE, the more rapidly $\beta_{\rm{EE}}^{*}(R_{\rm{tot}})$ decreases with the RSI power due to the same reason explained in the above. From the figure, the performance of FD communications largely depends on the strength of the RSI power.

\begin{figure}[!htp]
\center
\includegraphics[width=0.35\textwidth]{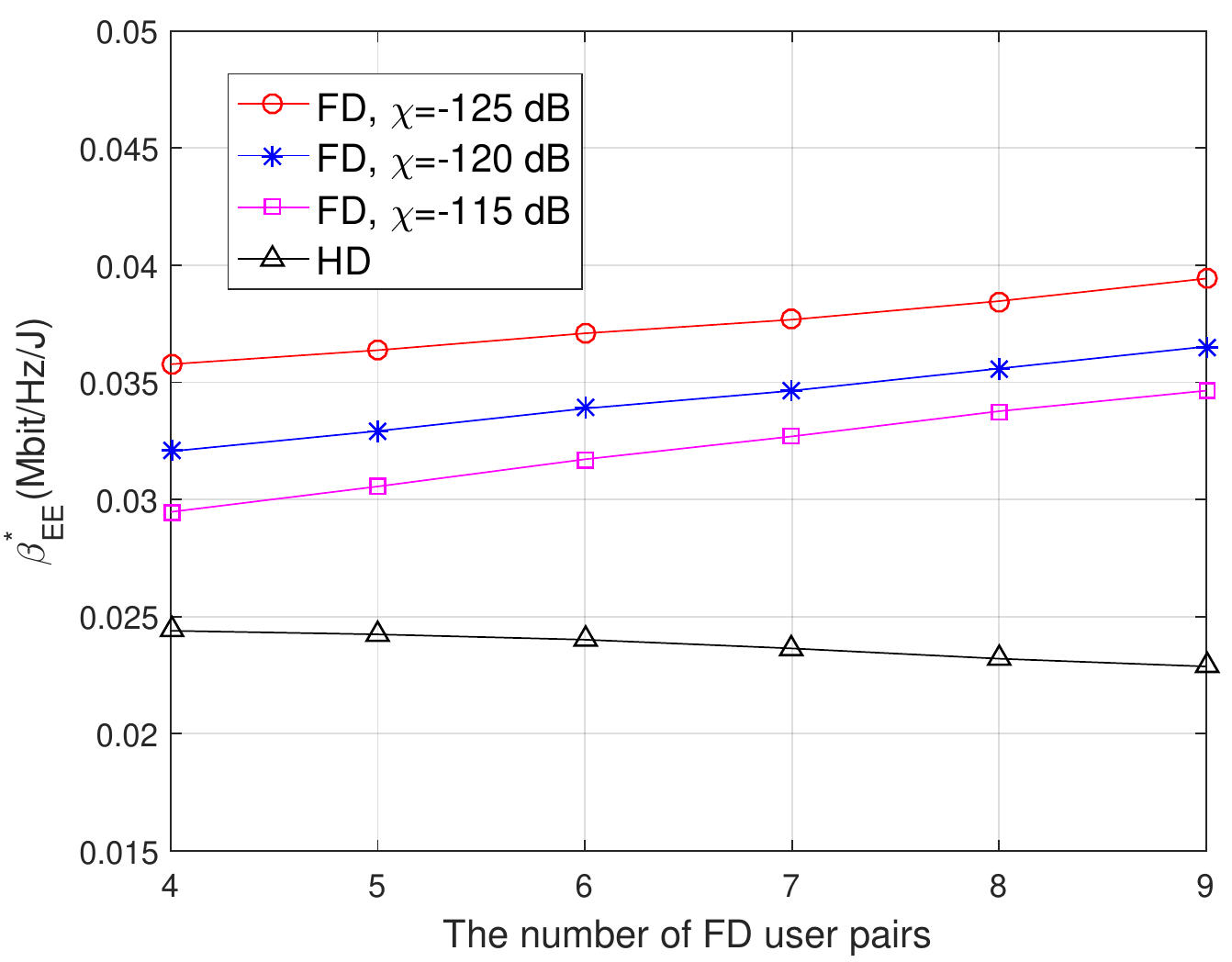}
\caption{EE-SE tradeoff with different numbers of user pairs. $\beta_{\rm{SE}}=8~{\rm{bit/s/Hz}}$.}\label{UE_chi}
\end{figure}

Fig. \ref{UE_chi} plots the maximum EE, $\beta_{\rm{EE}}^{*}$, for different numbers of user pairs, $M$. For the FD mode, it can be observed that $\beta_{\rm{EE}}^{*}$ increases with the number of FD user pairs because more users lead to more opportunities for paring users. However, for the HD network, $\beta_{\rm{EE}}^{*}$ decreases with the number of user pairs. This is because that more resource needs to be allocated to guarantee the fairness constraints in \eqref{eq:1.3} and \eqref{eq:1.4} as the number of users increases, leading to the degradation of EE.

\section{Conclusions}
In this paper, we have investigated the EE-SE tradeoff for FD enabled picocell networks. With fixed RSI power model, we first derive a necessary condition for FD communications to outperform HD communications in term of EE-SE tradeoff. Then, EE-SE tradeoff problem is solved in a closed-form way in the scenario of single pair of users. We also prove that, in this scenario, the EE is a quasi-concave function of the SE. We then extend the result into the multi-user scenario and also prove the quasi-concavity of the EE-SE relation. Based on this, the global optimal solution to maximize the EE for any given SE region is developed. Numerical results have verified the effectiveness of our analysis.

\begin{appendices}
\section{ }
To prove Theorem 1, we first consider the case where the data rate of FD mode is less than that of HD mode for a given transmit power, that is
\begin{equation}\label{eq:ap1}
R_{\rm{F}}\leq R_{\rm{H}}, \forall p_{ij}^{\rm{U}}+p_{ij}^{\rm{D}}=p_{\rm{H}}.
\end{equation}
Without loss of generality, we assume that $h_i^{\rm{U}}$ is no less than $h_j^{\rm{D}}$. Therefore, \eqref{eq:ap1} can be further expressed as
\begin{equation}
\dfrac{p_{ij}^{\rm{U}}h_i^{\rm{U}}}{1+\chi}+\dfrac{p_{ij}^{\rm{D}}h_j^{\rm{D}}}{1+p_{ij}^{\rm{U}}h_{ij}}+ \dfrac{p_{ij}^{\rm{U}}h_i^{\rm{U}}p_{ij}^{\rm{D}}h_j^{\rm{D}}}{(1+\chi)(1+p_{ij}^{\rm{U}}h_{ij})} \leq p_{\rm{H}}h_i^{\rm{U}}.
\end{equation}
Since we have $h_i^{\rm{U}}\geq h_j^{\rm{D}}$,
\begin{equation}\label{eq:ap2}
\dfrac{p_{ij}^{\rm{U}}h_i^{\rm{U}}}{1+\chi}+\dfrac{p_{ij}^{\rm{D}}h_i^{\rm{U}}}{1+p_{ij}^{\rm{U}}h_{ij}}+ \dfrac{p_{ij}^{\rm{U}}h_i^{\rm{U}}p_{ij}^{\rm{D}}h_j^{\rm{D}}}{(1+\chi)(1+p_{ij}^{\rm{U}}h_{ij})} \leq p_{\rm{H}}h_i^{\rm{U}},
\end{equation}
which is a sufficient condition for \eqref{eq:ap1}.

Moreover, by simplifying \eqref{eq:ap2}, a sufficient condition can be derived as
\begin{equation}\label{eq:ap3}
h_{ij}(1+\chi)\geq h_j^{\rm{D}}.
\end{equation}

Similarly, if $h_j^{\rm{D}}>h_i^{\rm{U}}$, a similar condition in \eqref{eq:ap3} by replacing $h_j^{\rm{D}}$ with $h_i^{\rm{U}}$ can be achieved. Therefore, the sufficient condition for \eqref{eq:ap1} can be given as
\begin{equation}\label{eq:ap4}
h_{ij}(1+\chi)\geq \min(h_i^{\rm{U}},h_j^{\rm{D}}).
\end{equation}


To prove Theorem 1, we assume that $R_{\rm{F}}=R_{\rm{H}}=R_{\rm{tot}}$ and the minimum transmit power of FD mode and HD mode are $p_{\rm{F}}$ and $p_{\rm{H}}$, respectively. Let $R_{\rm{H}}^{*}$ be the data rate of HD mode when the total transmit power is $p_{\rm{F}}$.

If \eqref{eq:ap4} is satisfied, we have $R_{\rm{H}}^*\geq R_{\rm{F}}=R_{\rm{H}}$. Since the data rate strictly increases with the transmit power of HD mode, we have $p_{\rm{F}}\geq p_{\rm{H}}$. This ends the proof.

\section{ }

Since $2^R$ and $\sqrt{2^R}$ are strictly convex on $R$ and affine transformation preserves convexity\cite{convex_optimization}, $P_1$, $P_2$, and $P_3$ are strictly convex on $R$. In the next, we prove the piecewise convex function in \eqref{eq:6} is also convex.

As we have mentioned before, the solution space of $(p_{ij}^{\rm{U}}, p_{ij}^{\rm{D}})$ is $C_1\cup (\overline{C_1}\& \overline{C_2})$ or $C_2\cup (\overline{C_1}\& \overline{C_2})$.
We first consider the case that $C_1\cup (\overline{C_1}\& \overline{C_2})$. In this case,
\begin{equation}
P_{ij}^{\min}=\left\{
\begin{split}
&P_1,~\text{if}~C_1,\\
&P_3, ~\text{if}~\overline{C_1}\& \overline{C_2}.
\end{split}
\right.
\end{equation}

Define $2^{R_e}=\frac{(h_j^{\rm{D}}-h_{ij})(1+\chi)}{h_i^{\rm{U}}-(1+\chi)h_{ij} }$ as the intersection of $P_1$ and $P_3$. The value and the first derivative of $P_{ij}^{\min}$ at $R_e$ are both  continuous, as
\begin{equation}\label{eq:ap11}
\left\{
\begin{split}
&P_1(R=R_e)=P_3(R=R_e)=(2^{R_e}-1)/h_j^{\rm{D}},\\
&P_1^{'}(R_e)=P_3^{'}(R_e)=2^{R_e}\ln2/h_j^{\rm{D}}.
\end{split}
\right.
\end{equation}

Without loss of generality, we assume $R_1<R_2$ for any $R_1,R_2\geq 0$. In the case that $R_1<R_2\leq R_e$ or $R_e\leq R_1<R_2$, we can easily derive that
\begin{equation}\label{eq:ap12}
P_{ij}^{\min}(R_2)-P_{ij}^{\min}(R_1)\geq {P_{ij}^{\min}}^{'}(R_1)(R_2-R_1),
\end{equation}
which is a necessary and sufficient condition for the convexity of $P_{ij}^{\min}$.

For the case that $R_1<R_e<R_2$, the left part of \eqref{eq:ap12} can be decomposed as
\begin{equation*}
\small
\left\{
\begin{split}
&P_{ij}^{\min}(R_2)-P_{ij}^{\min}(R_e)\geq P_1^{'}(R_e)(R_2-R_e)>P_1^{'}(R_1)(R_2-R_e),\\
&P_{ij}^{\min}(R_e)-P_{ij}^{\min}(R_1)\geq P_1^{'}(R_1)(R_e-R_1),
\end{split}
\right.
\end{equation*}
which indicates that the condition in \eqref{eq:ap12} is also satisfied in this case.

Therefore, it can be concluded that $P_{ij}^{\min}$ is convex when the solution space is $C_1\cup (\overline{C_1}\& \overline{C_2})$.
If the solution space is $C_2\cup (\overline{C_1}\& \overline{C_2})$, $P_{ij}^{\min}$ can also be proved as a convex function of $R$ in a similar way.


\section{  }

Define the super-level set of $\beta_{\rm{EE}}^{*}$ as $S_{\alpha}=\{R_{\rm{tot}}|\beta_{\rm{EE}}^{*}\geq \alpha \}$. $\beta_{\rm{EE}}^{*}(R_{\rm{tot}})$ is quasi-concave if $S_{\alpha}$ are convex for all $\alpha$.
\begin{itemize}
\item For $\alpha\leq 0$, $S_{\alpha}$ is its domain, which is obviously convex.

\item For $\alpha>0$, $S_{\alpha}=\{R_{\rm{tot}}|R_{\rm{tot}}-\alpha (\omega P_{ij}^{\min}(R_{\rm{tot}})+P_e)\geq 0\}$. Since $P_{ij}^{\min}(R_{\rm{tot}})$ is convex, $S_{\alpha}$ is convex.
\end{itemize}
This ends the proof.

\section{ }

Note that the constraints \eqref{eq:10.1}, \eqref{eq:10.2}, \eqref{eq:10.3}, and \eqref{eq:10.4} are linear. Therefore the problem in \eqref{eq:10} is convex if the objective function is convex. In the following, we prove that the objective function, $\sum_i\sum_j\gamma_{ij}P_{ij}$ is convex.

Define
\begin{equation}
\left\{
\begin{split}
&f_1(\gamma,R)=\gamma P_1(R/\gamma),\\
&f_2(\gamma,R)=\gamma P_2(R/\gamma), \\
&f_3(\gamma,R)=\gamma P_3(R/\gamma),
\end{split}
\right.
\end{equation}
where $P_i,(i=1,2,3)$ are defined in \eqref{eq:4-1}.
It can be easily prove that when $\gamma_{ij}\in(0,1]$, $f_1$, $f_2$, and $f_3$ are convex, since $\gamma2^{\frac{R}{\gamma}}$ and $\gamma\sqrt{2^{\frac{R}{\gamma}}}$ are convex.

Now we are ready to prove that $\sum_i\sum_j\gamma_{ij}P_{ij}$ is convex. Denote $f(\gamma_{ij},R_{ij})=\gamma_{ij}P_{ij}(\gamma_{ij},\hat{R}_{ij})$ in the sequel for simplicity. For notation simplicity, we use $\gamma$ as $\gamma_{ij}$ and $R$ as $\hat{R}_{ij}$.
As mentioned above, the solution space of $P_{ij}$ has two cases. For the first case where $C_1\cup(\overline{C_1}\&\overline{C_2})$, the corresponding $f$ can be expressed as
\begin{equation}
f(\gamma,R)=\left\{
\begin{split}
&f_1(\gamma,R),~\text{if}~2^{\frac{R}{\gamma}}\leq \frac{(h_j^{\rm{D}}-h_{ij})(1+\chi)}{h_i^{\rm{U}}-(1+\chi)h_{ij} },\\
&f_3(\gamma,R),~\text{otherwise}.
\end{split}
\right.
\end{equation}



Define ${\bf{x}}=(\gamma,R)$, $A_1=\frac{(h_j^{\rm{D}}-h_{ij})(1+\chi)}{h_i^{\rm{U}}-(1+\chi)h_{ij} }$, and $L=\big\{{\bf{x}}|2^{\frac{R}{\gamma}}=A_1\big\}$, and denote ${\bf{x_1}}=(\gamma_1,R_1)^T$ and ${\bf{x_2}}=(\gamma_2,R_2)^T$. Without loss of generality, we assume $0\leq\dfrac{R_1}{\gamma_1}<\dfrac{R_2}{\gamma_2}$. If $\dfrac{R_1}{\gamma_1}<\dfrac{R_2}{\gamma_2}\leq \log_2A_1$ or $\log_2A_1\leq \dfrac{R_1}{\gamma_1}<\dfrac{R_2}{\gamma_2}$, we can easily derived that
\begin{equation}\label{eq:ap13}
f({\bf{x_2}})-f({\bf{x_1}})\geq \nabla f({\bf{x_1}})^T({\bf{x_2}}-{\bf{x_1}}),
\end{equation}
which is the necessary and sufficient condition for the convexity of $f$.

Next, we will prove that if $\dfrac{R_1}{\gamma_1}<\log_2A_1<\dfrac{R_2}{\gamma_2}$, the condition in \eqref{eq:ap13} is also satisfied.
According to \eqref{eq:ap11}, for ${{\bf{x}}\in L}$, the first derivatives of $f_1$ and $f_3$ are equal, as
\begin{equation}
\left\{
\begin{split}
&\dfrac{\partial f_1}{\partial \gamma}\bigg|_{{\bf{x}}\in L}=\dfrac{\partial f_3}{\partial \gamma}\bigg|_{{\bf{x}}\in L}=\dfrac{2^{\frac{R}{\gamma}}(\gamma-R\ln2)-\gamma }{\gamma h_j^{\rm{D}}}\bigg|_{{\bf{x}}\in L},\\
&\dfrac{\partial f_1}{\partial R}\bigg|_{{\bf{x}}\in L}=\dfrac{\partial f_3}{\partial R}\bigg|_{{\bf{x}}\in L}=\dfrac{2^{\frac{R}{\gamma}}\ln2}{h_j^{\rm{D}}}\bigg|_{{\bf{x}}\in L}.
\end{split}
\right.
\end{equation}
Denote ${\bf{x_0}}=(\gamma_2,\gamma_2\log_2A_1)$. The left part of \eqref{eq:ap13} can be decomposed as
\begin{equation}\label{eq:ap14}
\left\{
\begin{split}
&f({\bf{x_2}})-f({\bf{x_0}})\geq \nabla f_1({\bf{x_0}})^T({\bf{x_2}}-{\bf{x_0}}),\\
&f({\bf{x_0}})-f({\bf{x_1}})\geq \nabla f_1({\bf{x_1}})^T({\bf{x_0}}-{\bf{x_1}}).
\end{split}
\right.
\end{equation}
Since $P_1$ is convex, it can be derived that
\begin{equation}
\begin{split}
&\nabla f_1({\bf{x_0}})^T({\bf{x_2}}-{\bf{x_0}})-\nabla f_1({\bf{x_1}})^T({\bf{x_2}}-{\bf{x_0}})\\
=&\big(P_1^{'}(\log_2A_1)-P_1^{'}(\dfrac{R_1}{\gamma_0})\big)\big(R_2-\gamma_2\log_2A_1\big)>0,
\end{split}
\end{equation}
which indicates that \eqref{eq:ap13} is satisfied.

%

For the second case where the solution space of $P_{ij}$ is $C_2\cup(\overline{C_1}\&\overline{C_2})$, it can also be proved that $f(\gamma_{ij},R_{ij})=\gamma_{ij}P_{ij}(\gamma_{ij},R_{ij})$ is convex.
Therefore, the objective function $\sum_i\sum_j\gamma_{ij}P_{ij}$ is convex.

\end{appendices}

\end{document}